\documentclass[twocolumn]{aastex62}
\usepackage[T1,T2A]{fontenc}
\usepackage[utf8]{inputenc}
\usepackage[russian,english]{babel}
\usepackage{comment}

\usepackage{natbib}
\usepackage{ulem,color,epstopdf} 
\usepackage{amsmath, graphicx}
\usepackage{courier}

\def\kw{{Konus-\textit{Wind}}}

\def\kws{{\textit{KW-Sun}}}

\begin{document}

\title{KW-Sun: The \kw\ Solar Flare Database in Hard X-ray and Soft Gamma-ray Ranges}

\author{A. L. Lysenko}
\affiliation{Ioffe Institute, Polytekhnicheskaya, 26, St. Petersburg, 194021 --  Russian Federation}
\email{alexandra.lysenko@mail.ioffe.ru}

\author{M. V. Ulanov}
\affiliation{Ioffe Institute, Polytekhnicheskaya, 26, St. Petersburg, 194021 --  Russian Federation}

\author{A. A. Kuznetsov}
\affiliation{Institute of Solar-Terrestrial Physics  (ISZF), Lermontov st., 126a, Irkutsk, 664033 --  Russian Federation}

\author{G. D. Fleishman} 
\affiliation{New Jersey Institute of Technology, University Heights, Newark, NJ 07102-1982 -- USA}

\author{D. D. Frederiks}
\affiliation{Ioffe Institute, Polytekhnicheskaya, 26, St. Petersburg, 194021 --  Russian Federation}

\author{L. K. Kashapova}
\affiliation{Institute of Solar-Terrestrial Physics  (ISZF), Lermontov st., 126a, Irkutsk, 664033 --  Russian Federation}

\author{Z. Ya. Sokolova}
\affiliation{Ioffe Institute, Polytekhnicheskaya, 26, St. Petersburg, 194021 --  Russian Federation}

\author{D. S. Svinkin}
\affiliation{Ioffe Institute, Polytekhnicheskaya, 26, St. Petersburg, 194021 --  Russian Federation}

\author{A. E. Tsvetkova}
\affiliation{Ioffe Institute, Polytekhnicheskaya, 26, St. Petersburg, 194021 --  Russian Federation}

\begin{abstract}

We present a database of solar flares registered by the \kw\ instrument during more than 27 years of operation, from 1994 November to now (2022 June).
The constantly updated database (hereafter \kws) contains over 1000 events detected in the instrument's triggered mode and is accessible online at \url{http://www.ioffe.ru/LEA/kwsun/}.
For each flare, the database provides time-resolved energy spectra in energy range from $\sim$20\,keV to $\sim$15\,MeV in FITS format along with count rate light curves in three wide energy bands G1 ($\sim$20--80\,keV), G2 ($\sim$80--300\,keV), and G3 ($\sim$300--1200\,keV) with high time resolution (down to 16\,ms) in ASCII and IDL SAV formats.
This article focuses on the instrument capabilities in the context of solar observations, the structure of the \kws\ data and their intended usage.
The presented homogeneous data set obtained in the broad energy range with the high temporal resolution during more than two full solar cycles is beneficial for both statistical and case studies as well as a source of context data for solar flare research.

\end{abstract}


\section{\label{sec_intro} Introduction}

Hard X-ray (HXR) and gamma-ray emissions play a primary role in the diagnostics of electron and ion acceleration in solar flares. 
They provide information about the spectrum, energetics and abundances of accelerated particles along with acceleration timescales \citep{Fletcher2011}.
Therefore hard X-ray/gamma-ray data with high temporal and energy resolutions are of particular interest.
The main mechanism responsible for HXR emission of solar flares is the bremsstrahlung emission of accelerated electrons \citep[see, e.g.,][and references therein]{Holman2011}.
Protons and heavier ions accelerated during a flare produce gamma-ray lines and a continuum due to numerous nuclear reactions \citep[see, e.g.,][and references therein]{Vilmer2011}.

Spectral, temporal, and spatial characteristics of HXR and gamma-ray emission in solar flares are very diverse. 
The range of flare energies emitted in HXR and gamma-ray ranges extends over many orders of magnitude, from $\sim$10$^{26}$\,erg \citep{Crosby1993} to $\sim$10$^{32}$\,erg for large eruptive flares \citep{Emslie2012}.
Energy spectra of the so-called microflares \citep[see, e.g.,][]{Hannah2008} barely reach ten keV, the majority of solar flares demonstrate emission up to several tens of keV, and in some flares the spectra extend to GeV energies \citep[see, e.g.,][]{Vestrand1999, Share2018}. 
The flare durations range from a fraction of a second for short impulsive bursts to hours for gradual flares \citep{Dennis1985, Crosby1993}. 

Despite several decades of research, many questions in solar flare physics are still under discussion and require further development of instrumentation and theory \citep{Fletcher2011}.
Among these questions are the particular mechanism (mechanisms) responsible for particle acceleration during reconnection; the cause of energy partitioning between different flare components including cases of 100\,\% efficiency of the electron acceleration \citep{Fleishman2022}; whether the same acceleration mechanism is involved in both short and long-duration HXR bursts, etc.

Due to the opacity of the Earth's atmosphere to X-rays and gamma-rays, the solar observations in these ranges became possible only with the beginning of the exoatmospheric astronomy.
Important results were obtained by low Earth orbit space observatories, among them are \textit{OSO-5} (1969--1983), \textit{OSO-7} (1971-1974), \textit{Hinotori} (1981-1991), \textit{Solar Maximum Mission} (\textit{SMM}, 1980-1989), \textit{Yokhoh} \citep[1991-2001]{Ogawara1991, Acton1992}, Reuven Ramaty High Energy Solar Spectroscopic Imager \citep[\textit{RHESSI},][2002-2018]{Lin2002}. 
Since 1975 space telescopes on board the Geostationary Operational Environmental Satellite (GOES) spacecraft series continuously monitor solar soft X-ray emission in two broad energy channels providing the commonly accepted X-ray classification of solar flares (classes A, B, C, M, and X) according to their power.
In 2020 February Solar Orbiter \citep[\textit{SolO},][]{Muller2020} was launched into a heliocentric orbit with the purpose of investigating solar and heliospheric physics using a payload of instruments designed for both remote and in situ sensing.
Among them is the Spectrometer/Telescope for Imaging X-rays \citep[\textit{STIX},][]{Krucker2020} which allows solar imaging in a 4--150\,keV range from an off-ecliptical positions.
Although now there are no solar-dedicated HXR and gamma-ray instruments operating in the near-Earth space, several high-energy detectors provide solar spectra and light curves in these ranges; e.~g.,  \textit{Fermi}-GBM \citep{Meegan2009} and \textit{Fermi}-LAT \citep{Atwood2009, Ajello2021} at low Earth orbit, which have operated since 2008, and \kw\ \citep{Aptekar1995}, which was launched in 1994 and operates in interplanetary space. 
Solar data from these instruments are freely accessible via the SolarSoft package\footnote{\url{https://www.lmsal.com/solarsoft/}}, Virtual Solar Observatory\footnote{ \url{https://sdac.virtualsolar.org/cgi/search}}, Interactive Multi-Instrument Database of Solar Flares and Helioportal\footnote{\url{https://sun.njit.edu/About/IMIDSFH.html}}, and other resources.

Here we present \kws , a database of \kw\ solar flare observations that cover more than two full solar cycles, from 1994 to nowadays. 
The main objectives of \kw\ are the studies of high-energy transient emission from distant astrophysical sources, such as cosmological gamma-ray bursts and soft gamma-repeater (magnetar) flares. 
Although the systematic use of the \kw\ data for solar physics applications began only recently, it has already yielded important results \citep[see a review by][]{Lysenko2020}.
The \kws\ database has been developing since 2016, and many improvements in data analysis and presentation have been made over the past few years.  

\kw\ has a number of advantages for solar flare studies compared to other instruments.
Among them are: high temporal resolution in the triggered mode, which allows one to reconstruct accelerated electron properties and obtain constraints on the acceleration mechanism \citep{Altyntsev2019, Glesener2018};  wide energy range that covers emission from accelerated electrons and a
good fraction of emission from nuclear reactions of accelerated ions \citep[eg.,][and references therein]{Lysenko2019}; and last but not least the instrument location in the interplanetary space that allows observations of the Sun  in stable background conditions during $\gtrsim$90\,\% of the time.  

\section{\label{sec_kw}\kw}

\begin{figure*}
  \begin{minipage}[b]{0.38\textwidth}
    \center{\includegraphics[scale=0.85]{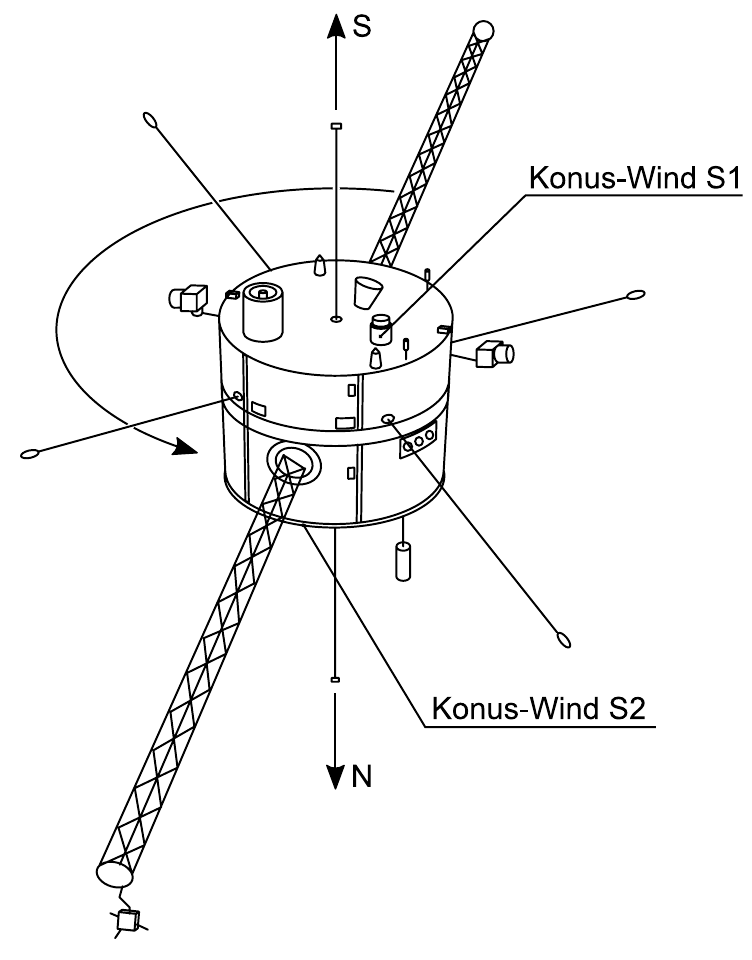} \\ (a)}
  \end{minipage}
  \hfill
  \begin{minipage}[b]{0.62\textwidth}
    \center{\includegraphics[scale=1.1]{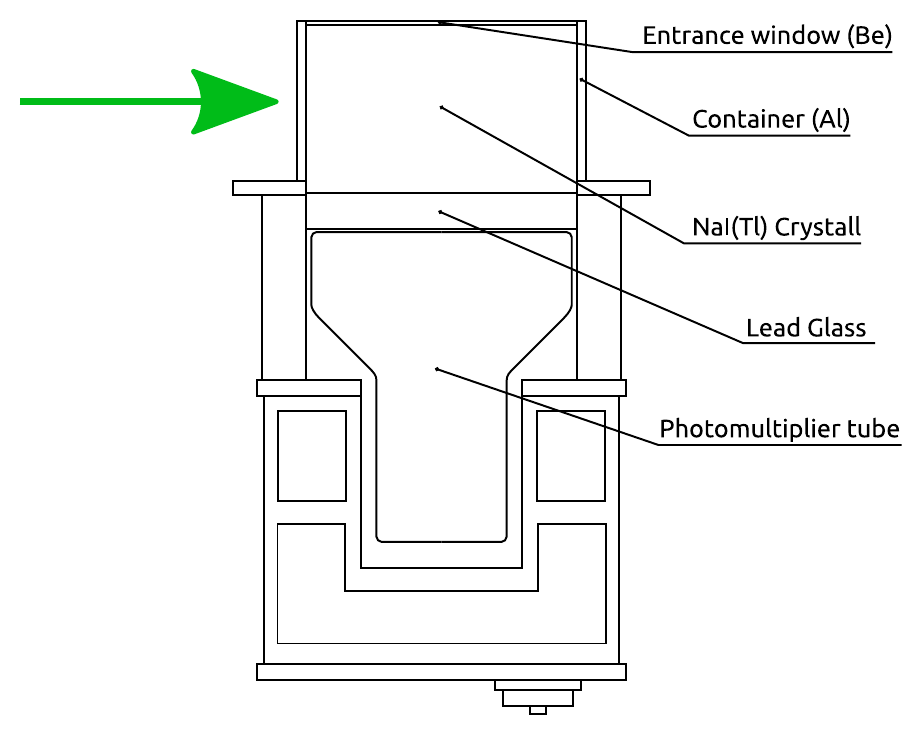} \\ (b)}
  \end{minipage}
\caption{\label{fig_konus} \kw : (a) the location of Konus detectors S1 and S2 on the spacecraft, ``N'' and ``S'' are the directions to the ecliptic poles and the circular arrow indicates the spin direction of the spacecraft; (b) a scheme of the Konus detector. The green arrow indicates the direction of solar emission.}
\end{figure*}

\begin{figure*}\centering
\includegraphics[width=0.49\textwidth]{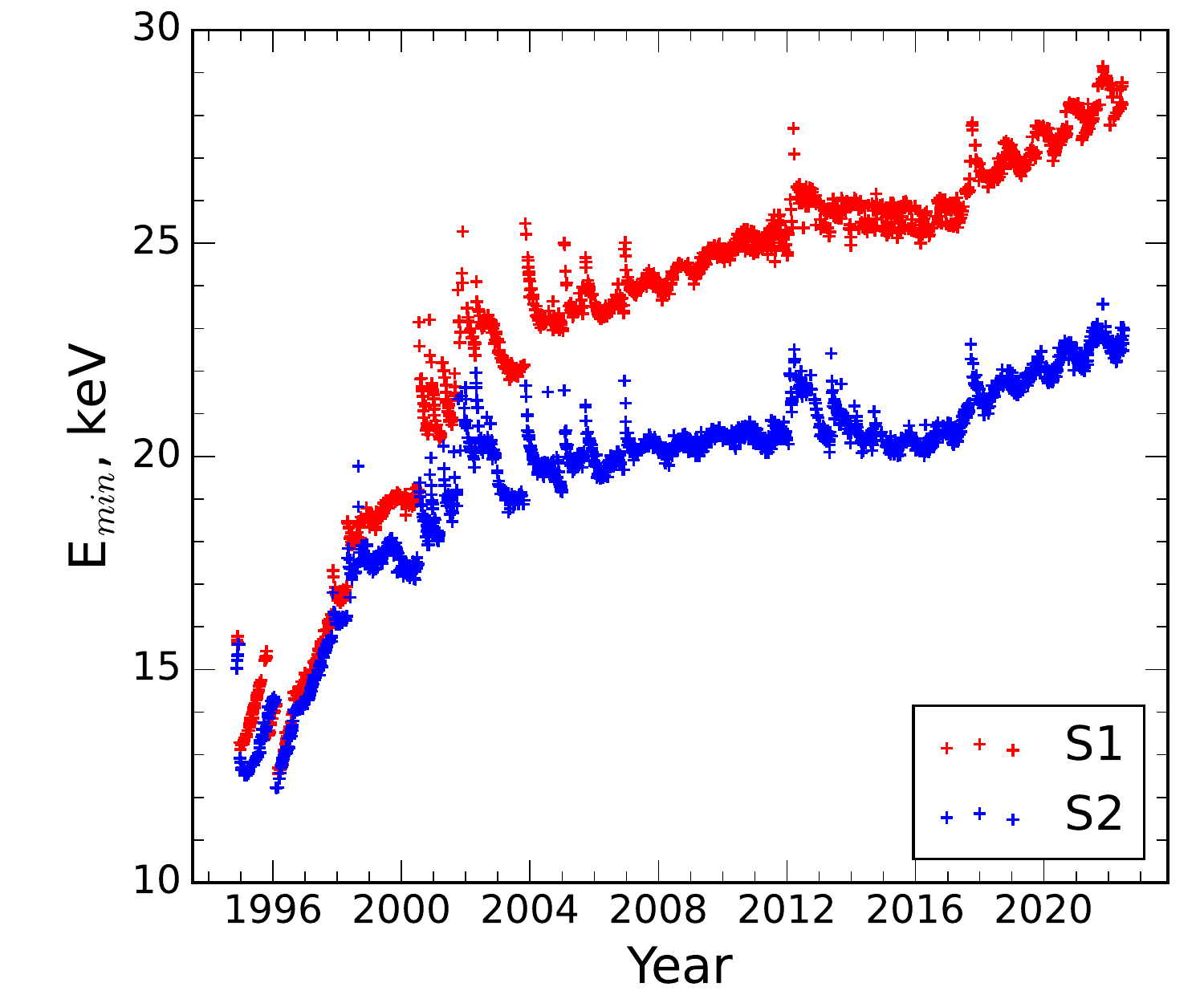}
\includegraphics[width=0.49\textwidth]{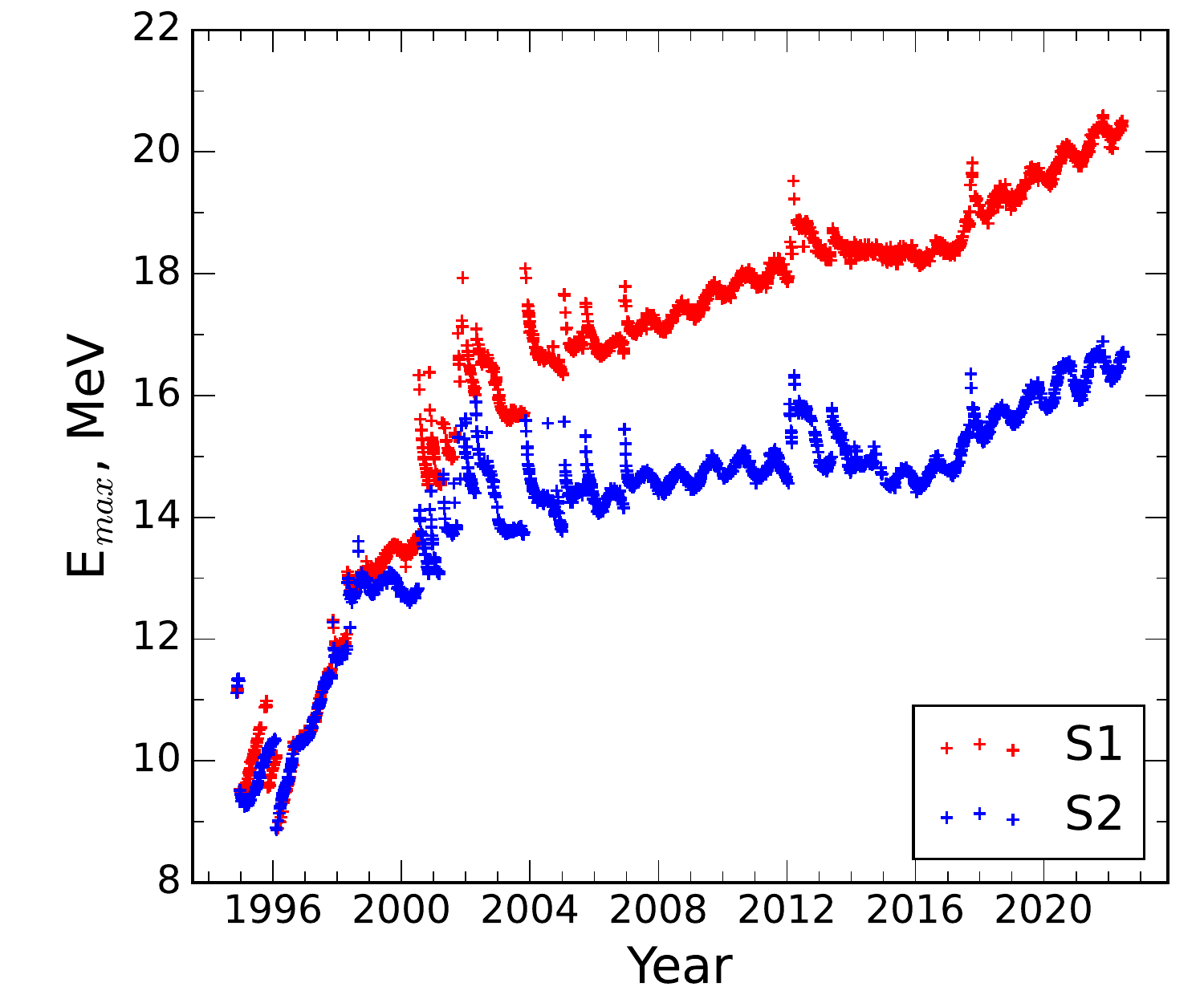}
\caption{\label{fig_calib} Temporal evolutions of \kw\ low-energy boundaries (left) and high-energy boundaries (right) for detectors S1 and S2.}
\end{figure*}

The \kw\ spectrometer was launched on board the NASA Wind spacecraft in 1994, November and operates till the present time \citep{Wilson2021}. 
\textit{Wind} is a spin stabilized spacecraft with spin axis aligned with ecliptic south and a spin period of $\sim$3 s.\footnote{The deviation of the spacecraft spin axis from the ecliptic north-south axis does not exceed 1$^{\circ}$ and the spin period varies gradually in the interval 3.0-3.2\,s.}
Since 2004, July, the spacecraft has been in orbit around Lagrange point L1 at $\sim$5 light seconds from the Earth.
Operating in the interplanetary space the instrument sees the Sun 24\,hours a day and being far from the Earth radiation belts it has an exceptionally stable background. 
The total fraction of time when the \kw\ data are unavailable or contaminated by a high solar particle background does not exceed 10\,\%.\footnote{Detailed information on the availability of \kw\ data for a particular time can be obtained from the authors on request.}

\kw\ consists of two cylindrical NaI(Tl) detectors S1 and S2 mounted on the opposite sides of the rotationally stabilized spacecraft so that they point to the southern and the northern ecliptical poles, respectively.
Consequently, the solar emission enters the detectors from the lateral side, at an incident angle of $(90 \pm1)^{\circ}$ to the axis, and is registered in both S1 and S2 (Figure~\ref{fig_konus}). Due to \textit{Wind} rotation, \kw\ suffers from occultations from other instruments and the spacecraft structures that is important for studies of flare time histories on millisecond timescales (see details in Section~\ref{ssec_lc}).

The instrument operates in two modes: waiting and triggered.
In the waiting mode, the count rates in three wide energy bands G1 ($\sim$20--80\,keV), G2 ($\sim$80--300\,keV), G3 ($\sim$300--1200\,keV) are available with time resolution of 2.944\,s
along with the count rate at energies $\gtrsim$10\,MeV (the Z channel). 
As the gamma-ray flux at the higher energies is low as compared to the charged particle background, the Z-channel data can be used to monitor flux variations of high-energy electrons and ions.

When the count rate in G2 exceeds a $\sim$9$\sigma$ threshold above the background on one of two fixed timescales, 1\,s or 140\,ms, the instrument switches into the triggered mode. 
Thus, rather spiky and spectrally hard (with a significant emission above $\sim$80\,keV) flares are recorded in the triggered mode, while more gradual and soft events are typically registered in the waiting mode only.

In the triggered mode, the count rate curves in G1, G2, and G3 are recorded with high time resolution during an interval of 229\,s. 
The time resolution constitutes 2\,ms for the time period from 0.512\,s before the trigger to 0.512\,s after the trigger time, 16\,ms from 0.512 to 33.280\,s after the trigger, 64\,ms from 33.280 to 98.816\,s after the trigger and 256\,ms for the remaining triggered time history. 
In the \kws\ database we merge adjacent 2\,ms bins into 16\,ms bins.
Along with the high resolution triggered mode light curves, the waiting mode record continues until $\sim$250\,s after the trigger time.

Starting from the trigger time, 64 energy spectra are accumulated, over pseudologarithmic energy scales, in two partially overlapping energy ranges PHA1 (63 channels, $\sim$20--1200\,keV) and PHA2 (60 channels, $\sim$0.35--15\,MeV).
For the first four spectra the accumulation time is fixed at 64\,ms and for the last eight spectra at 8.192\,s. 
For spectra from 5 to 56, the accumulation time varies between 256\,ms and 8.192\,s according to the count rate in G2: with the intensity increase the accumulation time decreases. 
Thus, the spectra may cover a time interval with the total duration varying from 79.104 s for very bright flares to 492 s for less intense events.
After the end of the trigger record, the measurements are stopped for $\sim$1\,hour due to the data readout, and only count rate in G2 is available with the time resolution of 3.68\,s. 

As the triggered mode record is limited in time ($\sim$250\,s for the time profiles and $\leq$492\,s for multichannel spectra) the record can stop before the actual end of the flare for long-duration events.

For each flare, the instrument energy scale is calibrated using the 1460\,keV line of $^{40}$K. 
Temporal evolution of energy boundaries for S1 and S2 is shown in Figure~\ref{fig_calib}.

The energy resolution (FWHM/E=$\Delta$E/E) of \kw\ depends on energy and constitute $\sim$20\,\% at 20\,keV and $\sim$5\,\% at 2\,MeV.
The spectral resolution of the detectors did not change significantly during the mission, and the corresponding resolution loss is less than a factor of 1.5 as compared to the ground-based calibrations \citep{Svinkin2016}.

For most flares, a standard \kw\ dead-time (DT) correction procedure (i.e., a simple non-paralyzable dead-time correction in each of the measurement bands, with a DT of a few microseconds for light curves and $\sim$42\,$\mu$s for multichannel spectra) provides a robust flux estimate. Details of DT corrections for very intense flares are given in Section~\ref{sssec_pileup}.

\section{\kws\ Database}

\begin{figure*}\centering
\includegraphics[width=1\textwidth]{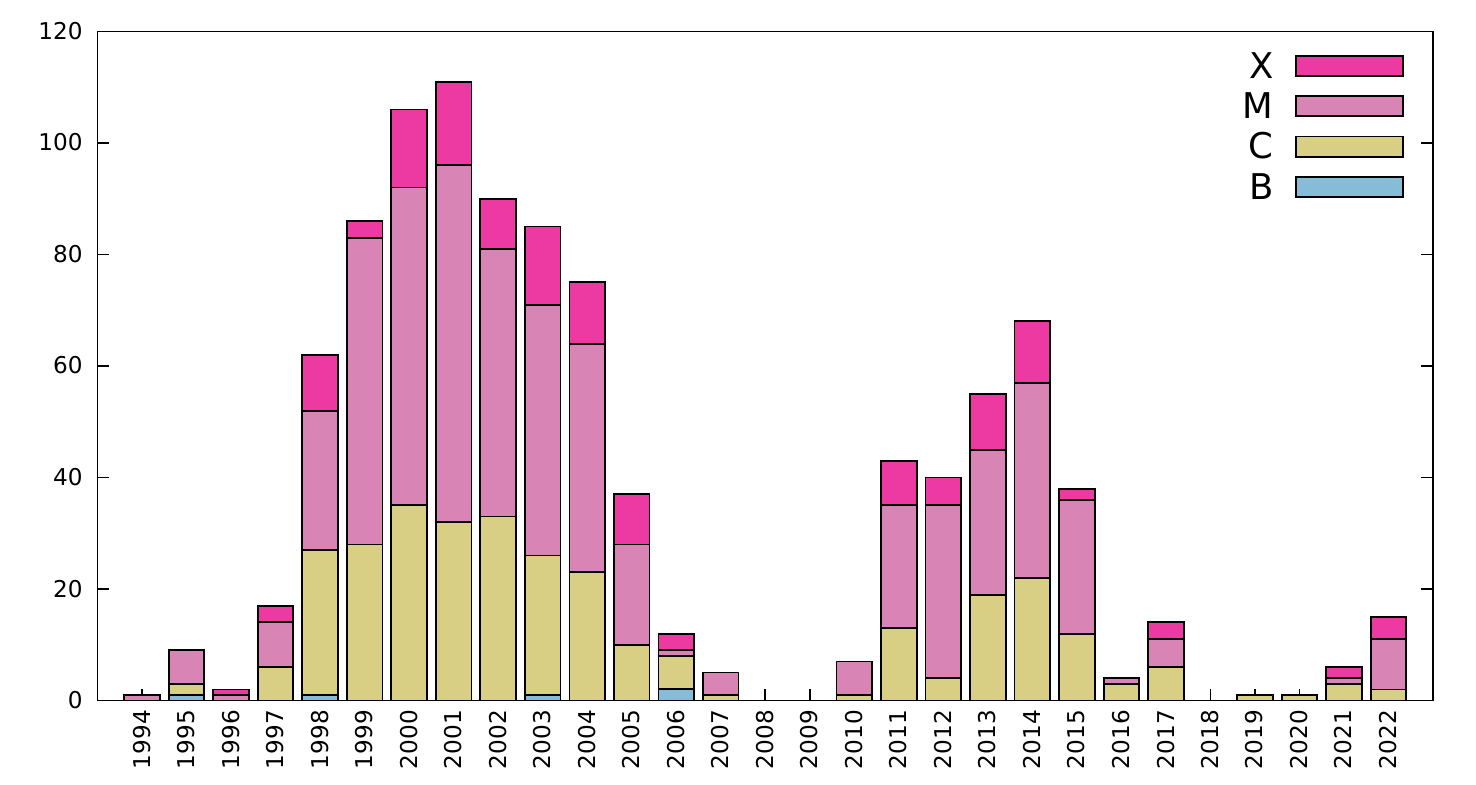}
\caption{\label{fig_stat} Annual rate of solar flares registered by \kw\ in the triggered mode categorized by the \textit{GOES} classes.}
\end{figure*}

\begin{figure*}\centering
\includegraphics[width=0.99\textwidth]{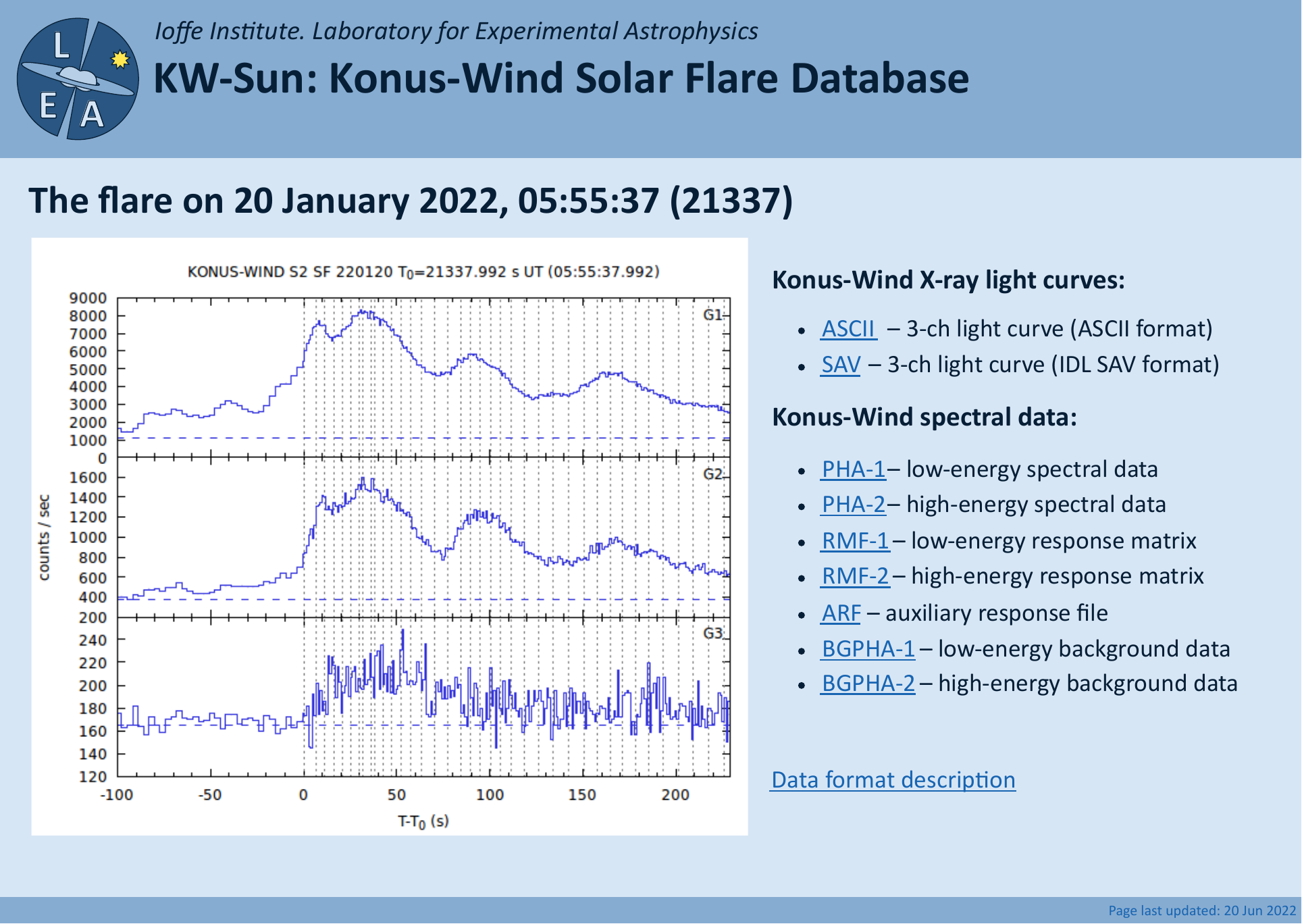}
\caption{\label{fig_screenshot} \kws\ web interface to the data for a specific flare, which displays the flare light curves in G1, G2, G3 and provides links to the data. 
Horizontal dashed lines represent background levels, and vertical dotted lines in the plot indicate time intervals of multichannel spectrum accumulation.}
\end{figure*}

By the time of writing, 2022 June, \kw\ registered $\sim$13,000 solar flares in the waiting mode, 1065 flares in the triggered mode, and, among them, 93 flares with emission at high, >1\,MeV, energies.
Annual statistics of the triggered solar flares categorized according to their \textit{GOES} classes is presented in Figure~\ref{fig_stat}. 

To identify a detected event as a solar flare we check if (i) the net count rates in both Konus detectors are close to each other, and, (ii) there is a corresponding flare in the \textit{GOES} event list\footnote{GOES event list is available via \url{https://hesperia.gsfc.nasa.gov/goes/}} or a simultaneous flux increase in the \textit{GOES} soft X-ray sensor data in the 1--8\,\AA\ and 0.5--4.0\,\AA\ bands. 
In controversial cases, we analyze spatial information from other telescopes. 

As \kw\ doesn't have an anticoincidence shield to filter out charged particles we use either count rate curve in the Z-channel or data from 3DP instrument \citep{Lin1995} located at \textit{Wind} spacecraft that provides information on the fluxes of  electrons in the 27--517\,keV energy band and protons in the 71--6800\,keV energy band. 
The total fraction of the \kw\ observation time contaminated by solar energetic particles is $\lesssim$8\,\%; data obtained during these intervals are not included to the database.

\begin{deluxetable*}{lllll}
\tablecolumns{5}
\tablewidth{0pc}
\tablecaption{\label{tab_kwsun} Data types presented in KW-Sun repository.
}
\tablehead{\colhead{Data type} & \colhead{Energy range}	& \colhead{Time resolution}	& \colhead{Record} & \colhead{Format} \\
\colhead{} & \colhead{}	& \colhead{}	& \colhead{duration\tablenotemark{a}} & \colhead{}
}
\startdata
Light curves & Three channels G1,G2,G3 & 16, 64, 256\,ms\tablenotemark{b}, & $\sim$250\,s & ASCII, IDL SAV \\
&  $\sim$20--1200\,keV & 2.944\,s\tablenotemark{c} & & \\
Multichannel & First range: 63 channels, & 64\,ms--8.192\,s & $\leq$492\,s & FITS, \textsc{OGIP} standard \\
spectra & 20-1200\,keV & & & PHA Type-II -- spectral files \\
 & Second range: 60 channels, & & & PHA Type-I -- background files\\
 & 350\,keV--15\,MeV & & & RMF, ARF -- response files  \\
\enddata
\tablenotetext{a}{After trigger time.}
\tablenotetext{b}{Triggered mode.}
\tablenotetext{c}{Waiting mode.}
\end{deluxetable*}

Currently, the \kws\ database contains light curves in the G1, G2, and G3 channels, multichannel spectra and detector response matrices for the flares registered in the triggered mode. 
The data are described in the following sections and summarized in Table~\ref{tab_kwsun}. 
A detailed data description is also available at the \kws\  website\footnote{\url{http://www.ioffe.ru/LEA/kwsun/kw-info.html}}. 
An example of a web interface to the data for a specific flare is given in Figure~\ref{fig_screenshot}.

As \textit{Wind} is located at distances up to $\sim$6 light seconds from Earth, for every flare in addition to the \kw\ trigger time (UT) we provide the trigger time corrected for the light propagation time from \textit{Wind} to the Earth center (geocenter time) -- in order to compare \kw\ observations with observations made by Earth-based or Earth-orbiting instruments.
It should be noted that the geocenter times may differ from UT at near-Earth spacecraft or ground-based locations by up to $\sim$20\,ms.

\subsection{\label{ssec_lc}Light Curves}

\begin{figure*}\centering
\includegraphics[width=0.49\textwidth]{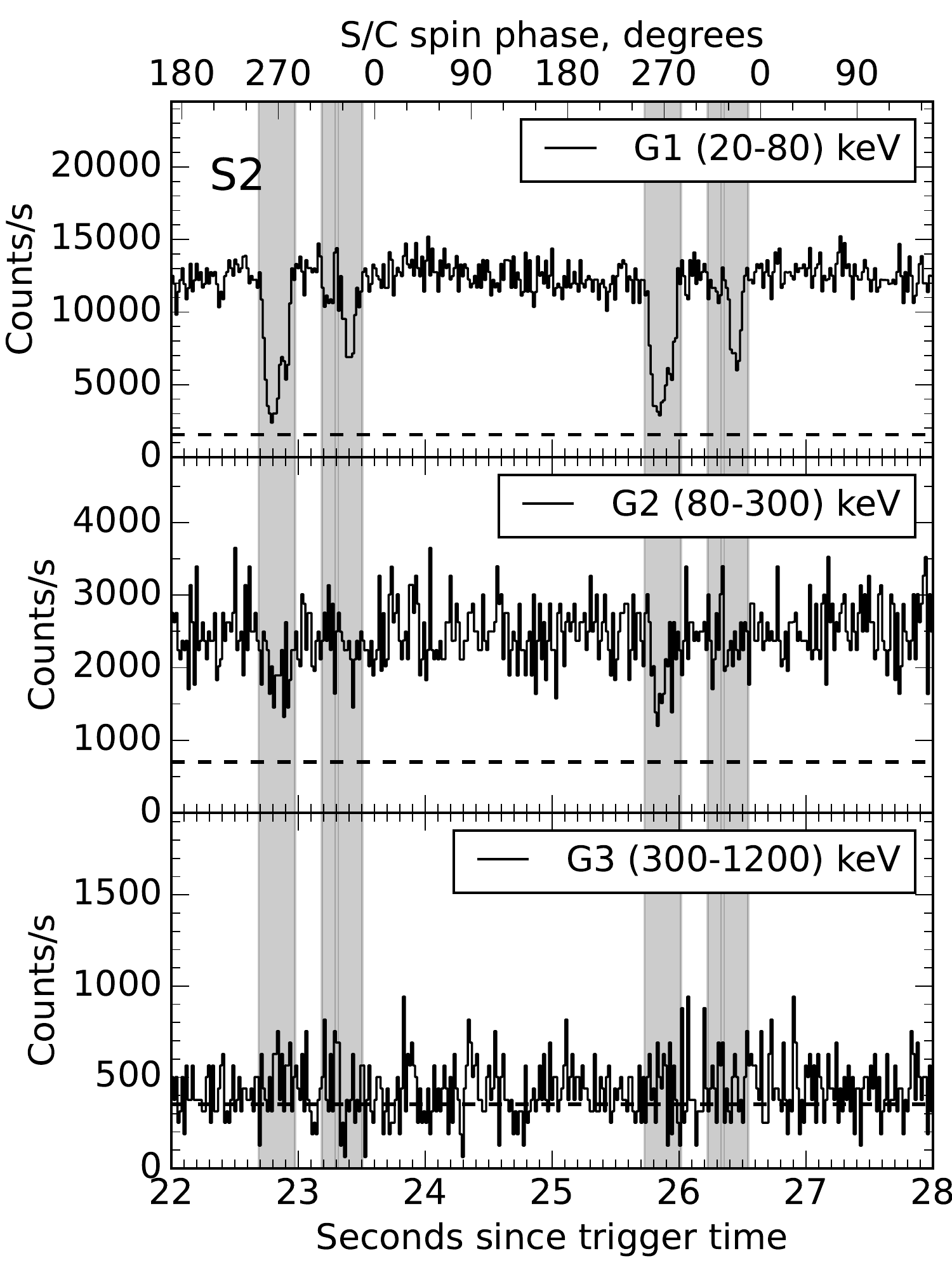}
\includegraphics[width=0.49\textwidth]{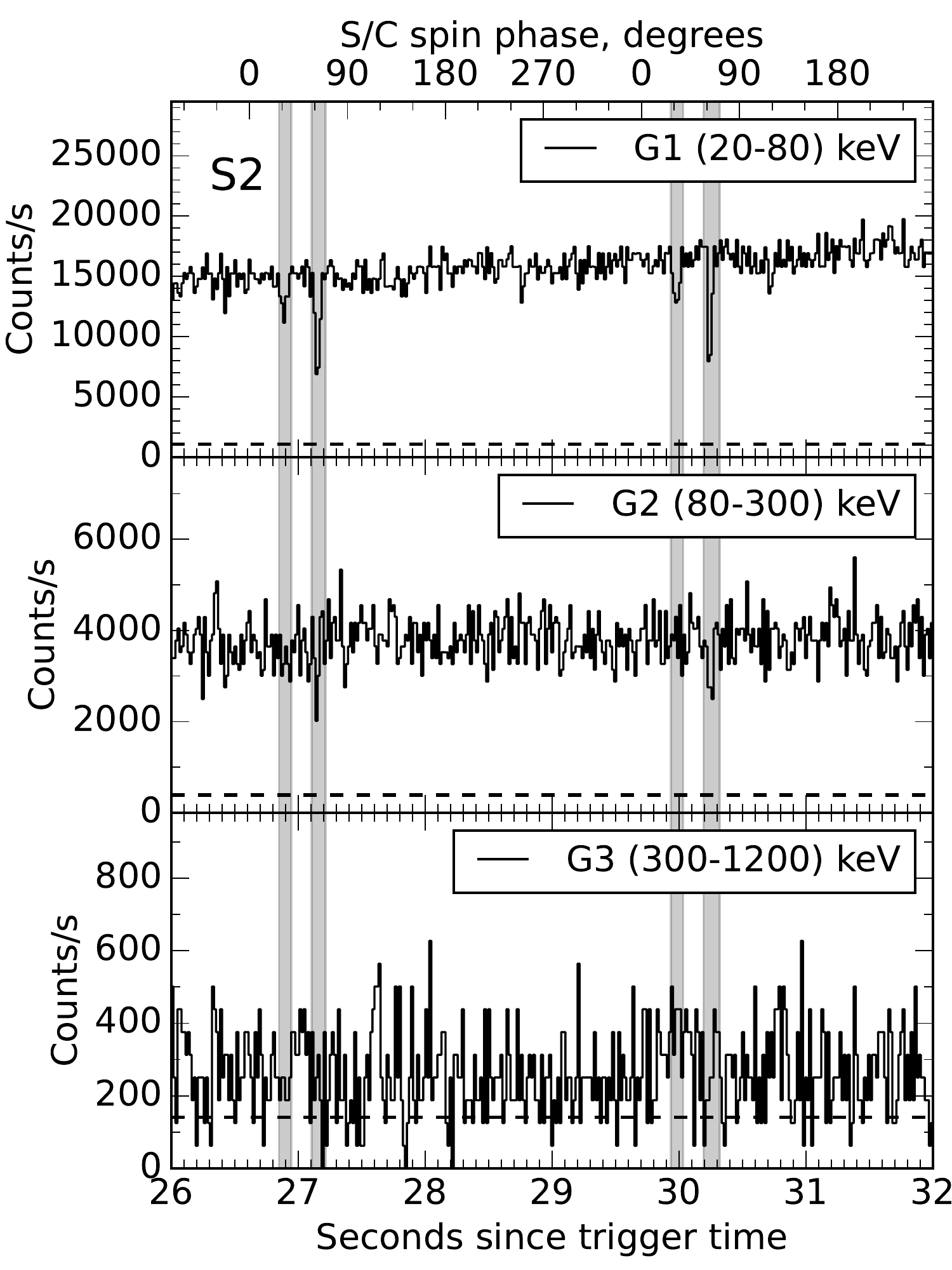}
\caption{\label{fig_occ_ex} Occultations in the \kw\ solar flare time profiles for S1 (left) and S2 (right) detectors. 
Gray shaded regions mark time intervals when the Sun was occulted partially or fully by other instruments or the spacecraft (S/C) structures. 
These intervals are removed from the \kws\ light curves.}
\end{figure*}

Light curves in the \kws\ database are available in two formats: ASCII and IDL SAV named as KWYYYYMMDD\_TSSSSS.txt and KWYYYYMMDD\_TSSSSS.sav respectively, where YYYYMMDD is the flare date and SSSSS is the trigger time in seconds UT.
The data description is available at the top of the ASCII file and in the ``description'' section of the SAV file. 

For each flare, the description contains the \kw\ trigger time ($t\_kw$ variable in the IDL SAV file), the corresponding geocenter time ($t\_earth$ variable in the IDL SAV file) and the calibrated energy boundaries of the G1, G2 and G3 channels ($channels$ variable in the IDL SAV file).

The light curve data are stored in 11 columns in the ASCII file and in four variables in the IDL SAV file: the start and end times of the time bin (columns $t1$ and $t2$ in the ASCII file and $tkw$ variable in the IDL SAV), relative to the trigger time; dead-time corrected count rates in G1, G2 and G3 (columns $G1$, $G2$, $G3$ in the ASCII file and $lc$ variable in the IDL SAV file); the background-subtracted and dead-time corrected count rates (columns $G1bgsub$, $G2bgsub$, $G3bgsub$ in the ASCII file and $lc\_bgsub$ variable in the IDL SAV file); and the count rate uncertanties at 1\,$\sigma$ level (columns $eG1$, $eG2$, $eG3$ in the ASCII file and $lc\_err$ variable in the IDL SAV file). 
The time history stored in the database includes several hundred seconds of waiting mode data measured before the trigger time, which, for gradually-rising flares, cover the beginning of the event (see Table~\ref{tab_kwsun}). 

\kw\ occultations by other instruments and the spacecraft parts appear on millisecond timescales as dips in the light curves with a period of $\sim$3\,s. 
Figure~\ref{fig_occ_ex} illustrates occultation influence on flare time profiles. 
The total durations of the dips per rotation period constitute $\sim$200\,ms for S2 and $\sim$600\,ms for S1. 
The occultation positions are calculated based on the spacecraft spin phase data, and the appropriate time intervals are removed from the light curves to ensure data reliability. 

An example plot of \kw\ time profile in G1, G2 and G3 is shown in Figure~\ref{fig_screenshot} for the M5.5 class flare of 2022-01-20. 

\subsection{\label{ssec_spec}Multichannel Energy Spectra}

For each flare, \kws\ provides multichannel energy spectra in the two energy ranges as PHA Type-II (PHA-II) FITS files, which contain multiple spectrum datasets. 
The corresponding background spectra (see below) are available as PHA Type-I (PHA-I) FITS files with a single dataset. 
The spectra are complemented with detector response matrices (in RMF format) and an ancillary response file (in ARF format). 
All these data conform \textsc{OGIP} specifications\footnote{\url{https://heasarc.gsfc.nasa.gov/docs/heasarc/ofwg/ofwg_intro.html}}. 
The detailed description of spectral and response files can be found in Appendix~\ref{app_ogip}.
The response matrices are calculated using the Geant4 toolkit \citep{Agostinelli2003}, for a parallel beam entering the detector at the incident angle of 90$^{\circ}$ to its axis. 
The energy channel boundaries of the matrix are adjusted to match the appropriate detector calibration.

Spectra for the first energy range ($\sim$20--1200\,keV) are measured by two redundant analyzers PHA1 and PHA3.
By default \kws\ contains data from PHA1 (the FITS files are named as KWYYYYMMDD\_TSSSSS\_1.pha), but in the cases of data gaps in PHA1, we provide the data from PHA3 (files named as KWYYYYMMDD\_TSSSSS\_3.pha).
Data for the second energy range ($\sim$0.35--15\,MeV) are stored in files named as KWYYYYMMDD\_TSSSSS\_2.pha (see Table~\ref{tab_kwsun}). 
Background spectrum files for the first energy range are named as KWYYYYMMDD\_TSSSSS\_1\_bg.pha (or KWYYYYMMDD\_TSSSSS\_3\_bg.pha if PHA3 is used, see above) and for the second energy range as KWYYYYMMDD\_TSSSSS\_2\_bg.pha. 
Response matrices for the first energy range are named KWYYYYMMDD\_TSSSSS\_1.rmf (KWYYYYMMDD\_TSSSSS\_3.rmf if PHA3 is used), for the second energy range -- KWYYYYMMDD\_TSSSSS\_2.rmf and an ancillary response file is named as KWYYYYMMDD\_TSSSSS.arf.

\begin{figure}\centering
\includegraphics[width=0.49\textwidth]{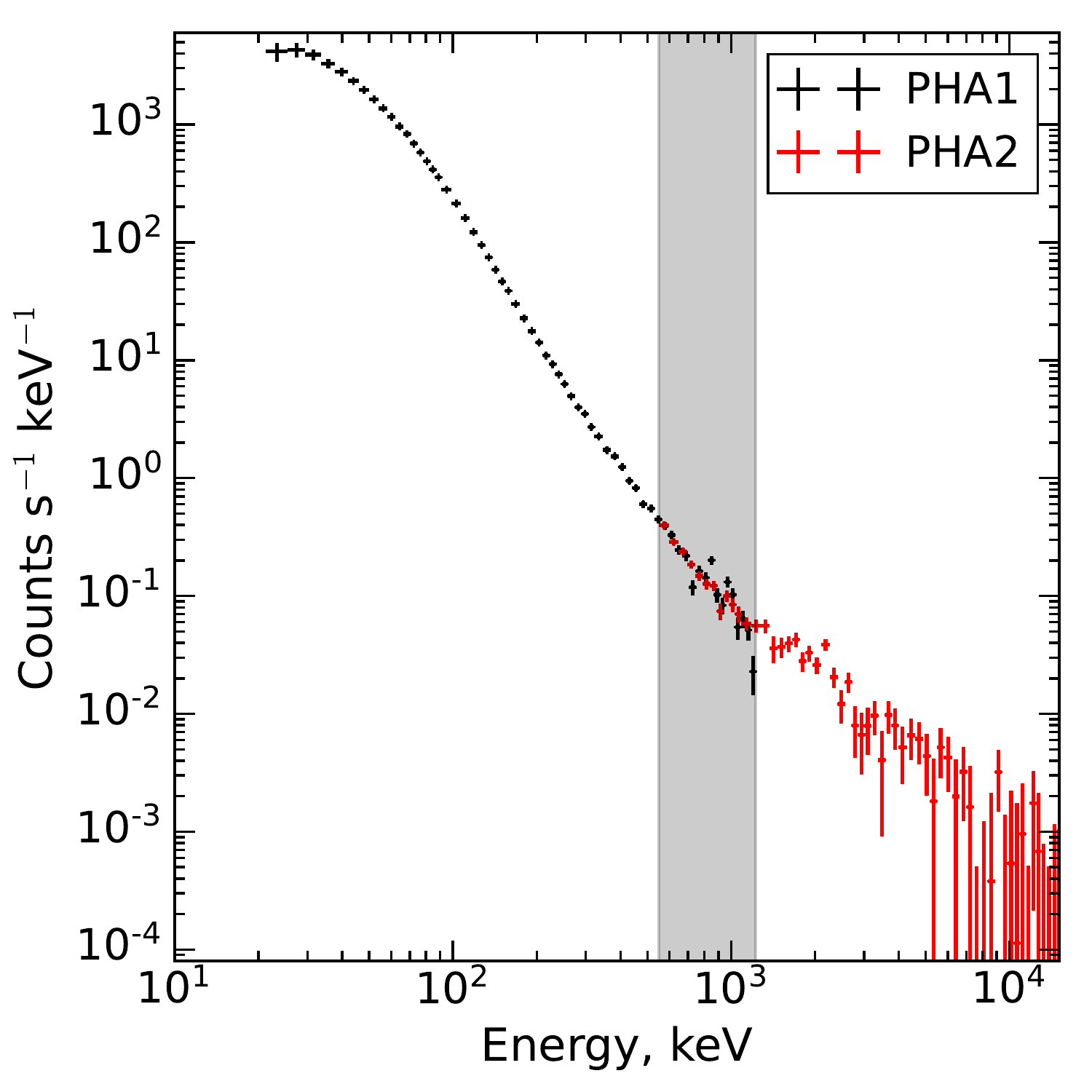}
\caption{\label{fig_ex_spec} Example of background-subtracted multichannel energy spectrum for the X9.3 class solar flare 2017-09-06 in the first energy range (PHA1, black) and the second energy range (PHA2, red). The gray shaded area marks the PHA1/PHA2 ovverlapping range.}
\end{figure}

An example of the multichannel spectrum in the first (PHA1) and the second (PHA2) energy ranges is plotted in Figure~\ref{fig_ex_spec} for the X9.3 class flare that occurred on 2017-09-06 and described in \cite{Lysenko2019}. 

For most solar flares ($\sim$90\,\%) data from the second, hard \kw\ energy range do not contain useful information on the emission. 
However, when data from both first (PHA1 or PHA3) and second (PHA2) ranges are used in the analysis, it is necessary to exclude the spectral channels of the first band in the overlapping range of energies (i.e. ignore channels 42-63 of PHA1 or PHA3) due to two reasons. 
First, in this range PHA1 (PHA3) and PHA2 measure the signals produced by the same detected photons (yet the PHA2 low-energy discriminator is set at a 10x higher level), and, second, for a typical steeply falling spectrum a smaller dead time fraction in PHA2 leads to better signal-to-noise ratio of the counts in the overlapping energy range (see Figure~\ref{fig_ex_spec}). 

For short-duration solar flares, that end before the accumulation of the multichannel spectra has finished, the background spectra are selected as measured at the end of the trigger record.
For the flares that last longer than the triggered record the background is taken from a nearby triggered event with similar background levels and close calibrations, and appropriate uncertanties are added in quadrature to the Poisson errors of the background counts (see Appendix. ~\ref{app_ogip})

A table containing time intervals covered by multichannel spectra for all flares registered in the triggered mode is available at \url{http://www.ioffe.ru/LEA/kwsun/KW_spectrum_times.txt}.

The \kws\ spectral data are suitable for analysis with standard tools for X-ray and gamma-ray spectral fitting, e.g., \textsc{XSPEC} \citep{Arnaud1996} and \textsc{OSPEX} \citep{Schwartz2002, Tolbert2020}, the latter is the part of the \textsc{SolarSoftWare} package.
We provide a Unix utility \textsc{SumKonusSpectra}\footnote{\url{http://www.ioffe.ru/LEA/kwsun/kw-info.html}} that extracts (and merges, if needed) individual \kw\ spectra from the PHA-II format files, which contain multiple multichannel spectra, to PHA-I format files which contain data on a single spectrum.
We recommend to use this utility when working with \textsc{XSPEC}.
The \textsc{OSPEX} package accepts input in PHA-II and allows searching for flares in the \kws\ database, downloading \kw\ spectral data automatically and analyzing them using its common routines. 
Both \textsc{XSPEC} and \textsc{OSPEX} load appropriate background and response files automatically when a spectral PHA file is selected for the analysis. 


\subsection{\label{ssec_high_loads}Instrumental effects at high intensities}

\subsubsection{\label{sssec_overflows}Integer overflows in low energy spectral channels}

At high incident photon fluxes, integer overflows in low-energy channels of PHA1 (PHA3) ($\leq$40\,keV) may occur that result in underestimation of the total count number, DT, and flux; and also in the spectrum distortion.  
We use a special algorithm for the overflow corrections, but, for some very intense flares, incorrect values in the first channels of  PHA1 (PHA3) can appear anyway.
Thus, in the cases of bad fit residuals in the low-energy channels, it is recommended to exclude these channels from the analysis. 

\subsubsection{\label{sssec_pileup}Circuit saturation and pulse pileup effects\footnote{The detailed description of pile-up corrections for \kw\ data is given in Appendix of \cite{Lysenko2019}.}}

For intense events, the standard \kw\ dead-time correction procedure (see Section~\ref{sec_kw}) results in flux underestimation due to saturations in the instrument's signal shaper and counter logic circuits. 
Such saturations are not negligible in PHA1 (PHA3) at incident photon rates per detector area $n$ exceeding $\sim$2$\times$10$^4$\,photon\,s$^{-1}$, and in the G1, G2, and G3 channel counters at $n\gtrsim 2.5 \times$10$^5$\,photon\,s$^{-1}$.
At considerably higher photon rates, the shape of the spectrum can also be distorted due to the pulse pileup effect. 
We studied the influence of the pileup effect using Monte-Carlo modeling and found that it becomes significant for PHA1 (PHA3) at $n\gtrsim 5 \times$10$^4$\,photon\,s$^{-1}$. 

For the events of moderate intensity (2$\times$10$^{4}$\,photon\,s$^{-1}$ < $n$ < 5 $\times$10$^4$\,photon\,s$^{-1}$, $\sim$15\,\% of all triggered flares) the flux in PHA1 (PHA3) can be underestimated, but this uncertainty does not exceed $\sim$10\,\%, thus no additional corrections are needed.
For high intensity events ($n$\,>\,5$\times$10$^4$\,photon\,s$^{-1}$, $\sim$8\,\% of all triggered flares) the corrections are applied for both spectral shape and flux.
In practice, the saturation correction to PHA1 (PHA3) count rate can be estimated using dead-time corrected counts in the overlapping PHA1/PHA2 energy range (see Section~\ref{ssec_spec}).
For the cases of additional dead-time corrections applied for high intensity flares two columns with count and exposure uncertanties are added to the PHA files. 
These systematic errors are taken into account automatically both in \textsc{XSPEC} and \textsc{OSPEX} (see Appendix~\ref{app_ogip}).
As the pileup correction procedure is time consuming, corrected PHA1 (PHA3) data are available only for a subset of intense flares, while for other flares spectral data for the 1st energy range are not available from the database and the appropriate note is added to the web page of the flare. In the latter case, we encourage the user to contact the \kws\ team by e-mail for pileup-corrected spectra of the event they are interested in.

\subsubsection{Uncertanties with energy calibration}
One more issue for high-intensity events is the calibration uncertainties resulting from the fact that the lines used in the calibration are faintly distinguishable above the continuum.
Although these uncertainties do not exceed a few per cent, they may be important in the gamma-ray emission line studies.

\section{Summary and Conclusion}
In this paper, we presented \kws\ -- the online catalog of solar flares registered by \kw\ experiment in hard X-ray and soft gamma-ray ranges during more than two full solar cycles. 
The database contains light curves with high (down to 16\,ms) time resolution and multichannel spectra in the wide energy range ($\sim$20\,keV--15\,MeV) for more than a thousand solar flares detected by the instrument in the triggered mode. 

The main area of application of the \kw\ data for Solar physics is the study of the emission from accelerated electrons and ions.
The high time resolution in the triggered mode allows to examine properties of nonthermal electrons on a subsecond timescales which is crucial for the research of the acceleration mechanisms during magnetic reconnection. 
Other advantages of \kw\ for solar flare stuies are the absence of the Earth occultations and extremely stable background conditions due to its location in the interplanetary space. 
The presented homogeneous data set can be used for both statistical and case studies of solar flares. 
Now with the beginning of a new solar cycle, this may be especially in demand as there is no solar-dedicated instrument in hard X-ray range near the Earth.

\kws\ database is a work in progress, data on new flares are added as they become available. 
The major ``to do task'' is to extend \kws\ with numerous solar flares registered in the waiting mode and to provide  response matrices for 3-channel time history data, which will allow spectral fitting of \kw\ G1, G2, and G3 data with simple spectral models at fine time scales. 
Future enhancements of \kws\ will be available via the database web site at Ioffe\footnote{\url{http://www.ioffe.ru/LEA/kwsun/}}.

\acknowledgements
We thank Prof. A. K. Tolbert for her help with incorporating \kw\ data in \textsc{OSPEX} package. 
G.D.F. was supported in part by NSF grant AGS-2121632 and NASA grant 80NSSC19K0068 to New Jersey Institute of Technology.

\appendix

\section{\label{app_ogip} Structure of FITS files in \textsc{OGIP} stardard}

PHA-II files contain EBOUNDS extension with boundaries of energy channels (columns E\_MIN and E\_MAX) and SPECTRUM extension which includes start (column TSTART), stop (column TSTOP) and exposure (column EXPOSURE) time of spectrum accumulation and $N_{sp}$$\times$$N_{ch}$ matrix (COUNTS) with count number in each of $N_{ch}$ channels for $N_{sp}$=64 multichannel spectra. 
The separate file with background multichannel spectra in PHA-I format contains EBOUNDS extension with energy boundaries equal to those of a flare spectrum and a single spectrum in SPECTRUM extension with columns CHANNEL, RATE, STAT\_ERR, SYS\_ERR, GROUPING and QUALITY. 

Standard \kw\ dead time corrections (Section~\ref{sec_kw}) are taken into account in the EXPOSURE column for PHA-II files and in the RATE and STAT\_ERR columns of PHA-I files. 
Systematic errors associated with additional dead-time corrections for very intense events (see Section~\ref{sssec_pileup}) are added in quadrature to the statistical errors in each channel, and the resulting errors are written to the STAT\_ERR column in the PHA files. 
This column (if exists) is automatically used by XSPEC and OSPEX instead of Poisson errors for count spectra. 
Exposure errors caused by these corrections are added, for the information, to the EXPOSURE\_ERR column in the PHA file. 

RMF files consist of EBOUNDS extension and MATRIX extension. 
MATRIX includes columns ENERG\_LO and ENERG\_HI with $N_i$=263 incident energies ${E_i}$ of test quanta used for response modeling and $N_i$$\times$$N_{ch}$ matrix with probabilities of registering gamma-quantum of energy $E_i$ in channel $N_i$ compressed by a special algorithm\footnote{Response matrix compression is described in \textsc{OGIP} standard.}. 
ARF files contain SPECRESP extension with effective areas calculated for $N_i$=263 incident energies ${E_i}$ of test quanta.

\clearpage

\bibliography{KWSun}

\end{document}